\RequirePackage{tikz} 
\documentclass[nofootinbib,twocolumn, superscriptaddress,secnumarabic,amssymb, aps,prb,showpacs,10pt,floatfix]{revtex4-2}
\usepackage{graphicx}
\usepackage{dcolumn}
\usepackage{bm}
\usepackage{color}
\usepackage{amsmath}
\usepackage{amssymb}

\usepackage{tikz}
\usetikzlibrary{arrows,shapes}
\usetikzlibrary{arrows.meta}
\tikzset{%
  >={Latex[width=2mm,length=2mm]},
            base/.style = {rectangle, rounded corners, draw=black,
                           minimum width=1.cm, minimum height=0.8cm,
                           text centered, font=\sffamily},
  activityStarts/.style = {base, fill=green!60},
       quantity/.style = {base, fill=red!60},
    activityRuns/.style = {base, fill=green!30},
         process/.style = {base, minimum width=1.cm, fill=blue!60,
                           font=\ttfamily},
}
\pgfdeclarelayer{background}
\pgfdeclarelayer{foreground}
\pgfsetlayers{background,main,foreground}

\definecolor{gray}{gray}{0.5}

\newcommand{\SF}{S_{\scriptscriptstyle\mathrm F}}
\newcommand{\Vph}{V_{\!\scriptscriptstyle\mathrm{ph}\!}}
\newcommand{\Vqh}{\widetilde V_{\!\scriptscriptstyle\mathrm{ph}\!}}

\newcommand{\wI}{w_{\mathrm{I}\!}}

\begin{document}


\title{Optimized correlations inspired by perturbation theory}

\author{Martin Panholzer}
\affiliation{Institute for Theoretical Physics, Johannes Kepler University Linz, Austria}
\affiliation{European Theoretical Spectroscopy Facility (ETSF)}
\author{Raphael Hobbiger}
\affiliation{Institute for Theoretical Physics, Johannes Kepler University Linz, Austria}
\author{Helga B\"ohm}
\affiliation{Institute for Theoretical Physics, Johannes Kepler University Linz, Austria}

\email{martin.panholzer@jku.at}



\date{\today}

\begin{abstract}
We study the accuracy of analytical wave function based many-body methods derived by energy minimization of a Jastrow-Feenberg ansatz for electrons (`Fermi hypernetted chain / Euler Lagrange' approach). Approximations to avoid the complexity of the fermion problem are chosen to parallel successful boson theories and be computationally efficient. For the three-dimensional homogeneous electron gas, we calculate the correlation energy, the pair distribution function and the static structure function in comparison with simulation results. We also present a new variant of theory,  
which is interpreted as approximate, self-consistent sum of ladder and ring diagrams of perturbation theory.  
The theory performs particularly well in the highly dilute density regime. 

\end{abstract}

\maketitle


\section{Introduction \label{sec: Intro}}

The quantum many-body (MB) problem, a numerically hard problem, cannot be solved straightforwardly. A potent approximation strategy requires to thoroughly understand, for each particular system, how the relevant physics manifests itself in the existing competing theories. Here, we combine experience from three prospering fields, quantum Monte Carlo (MC), perturbation theory (PT), and variational Jastrow-Feenberg (JF) approaches \cite{Triest2002}.
We aim at a better comprehension of where, depending on the system parameters, specific approximations perform better. Our study on fermions with an only distance-dependent interaction \(v(r)\) can be suitably generalized to dipoles \cite{Macia2012}, mixtures \cite{StMZ18:selfbound, KSSZ98:concentrationdependence}, or lattice models \cite{Wang1990}. The homogeneous electron gas (HEG) is of particular interest because it is i) extremely relevant for electronic structure methods, ii) well studied with various methods in its entire density range, and iii) its random phase approximation (RPA) is well defined.

Monte Carlo methods 
yield benchmark results for the ground state energy \cite{Ceperley1980}, the pair distribution function \cite{Gori-Giorgi2000,Spink2013} or the static density-density response function \cite{Moroni1995}. 
Extending the applicability to excited states is an active field with utmost numerical demands \cite{Motta2015,Nava2013}. 

Based on Feynman diagrams, PT is proven systematic treatment of MB systems, widely used for calculating both, excited state \cite{Rohlfing2000,Onida2002} and ground state properties \cite{Holm2000,Maggio2016,Bechstedt2018}. Practical implementations allow to retain specific, appropriately chosen classes of diagrams. A prime example is the importance of self-energy graphs for quasiparticles, successfully hand\-led in `\(GW\) summations' \cite{Aryasetiawan1998, Onida2002}. Another class arises from the Bethe-Salpeter equation (BSE), relevant for correct results on exciton binding observed by optical absorption \cite{Albrecht1998,Rohlfing2000,Gatti2013}. 

Wave function based methods include physical intuition right from the start, either via a parametrized or a general functional form 
The Jastrow-Slater wave function, $\psi=F \phi_{\scriptscriptstyle0\,}$, works excellently; for homogeneous cases $\phi_{\scriptscriptstyle0}$ is a plane wave Slater determinant, and \(F\!=\>\){\small\(\prod_{i<j\,}\)}\(\exp{\!\big(\frac12 u_2(|r_i\!-\!r_j|)\big)}\) accounts for correlations\footnote{The bosonic JF form is exact, when correlations \(u_{n}\) are included to all orders.}. The approach attains its full power when \(u_2(r)\) is optimally determined via functional variation. Inspired by cluster expansions developed for classical liquids \cite{Hansen2013},  
the pair distribution function $g(r)$ is here explored in the Hyper Netted Chain formalism, termed FHNC in its fermion version \cite{Krotscheck2000, Polls2002, Kallio1996}. 

The \textit{full} diagrammatic formalism exactly maps $u_2$ to all observables
as illustrated in fig.\,\ref{fig:FHNC} (left). Graphs are classified by their topological structure as nodal, non-nodal, or elementary diagrams. 
The latter are arbitrarily difficult (similar to PT, where vertex corrections are complicated).
Energy minimization determines the best $g$ via a corresponding Euler-Lagrange (EL) equation%
\footnote{In the literature, `(F)HNC' may denote both, the exact cluster expansion with \emph{all} elementary diagrams, and `(F)HNC/0', where these are omitted. Functional optimization is emphasized by `-EL', e.g. `(F)HNC/0-EL'.}
(fig.\,\ref{fig:FHNC}, right).
An exact result obtained this way would essentially equal that of variational MC (VMC) with a parametrized $u_2$ allowing to reach the functional result.

The theory, as any, relies on approximations. Although not reaching quite the MC accuracy, FHNC is numerically by orders of magnitude less demanding\footnote{For homogeneous systems FHNC/0-EL scales as $\,n \ln n\,$ for $n$ sampling points of $g(r)$.}. 
It thus allows a highly efficient evaluation of observables in a multi-variable space, depending, e.g., on position \(r\), density \(\rho\), spin \(\sigma\), and valley index \cite{GSV06:valley}. Most important, the FHNC can be systematically improved by topping it with PT, termed `correlated basis functions' approach (CBF) \cite{BKro02}. For the ground state, this is comparable with stepping from VMC to released node diffusion MC. The CBF route is also the prime tool to extend the theory to excited states \cite{nature, Bohm2010, Panholzer2009, Panholzer2018, Astrakharchik2016, Akaturk2018}. Again, in terms of accuracy versus computation time, obtaining dynamic properties via FHNC+CBF is much more efficient than by MC \cite{Nava2013, Astrakharchik2016}. 

At first sight, PT and optimized JF approaches appear very different. Their link was demonstrated for bosons by Jackson et al.\ \cite{Jackson1982, Jackson1985}, who showed which approximation to the sum of parquet diagrams leads to the HNC equations. 
We here extend their studies to fermions. In particular, we derive an approximation to the particle-particle ladder, inherent to the variational Jastrow-Slater ansatz and identical to the boson ladder equation corrected by a ``potential'' constructed from the Slater-determinant's density matrix.
Accounting for self-consistently summed ladder and ring diagrams, it is excellent for highly dilute HEGs ($r_s\!>\!5$; the density \(\rho\) is \(3/4\pi(a_{\scriptscriptstyle0}r_s)^3\), and \(a_{\scriptscriptstyle0}\) the effective Bohr radius). 

We further demonstrate that this and two other FHNC variants \cite{Krotscheck2000, Kallio1996} perform very well even in their most basic, boson-like versions. All are easier to implement than the more often employed approach of Singwi et al.\,\cite{Singwi1969} (STLS) and give better results. Additionally, they yield effective interactions for PT, providing physical insight by their connection to diagrammatic (sub-)classes.


\tikzstyle{fancytitle} =[fill=none, text=black!70,font=\bf]
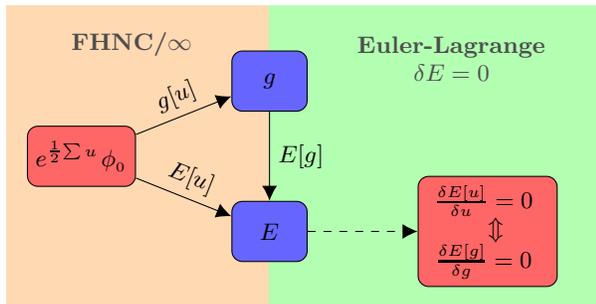
\begin{figure}
\centering
\begin{tikzpicture}[node distance=1.5cm,
    every node/.style={fill=white, font=\sffamily}, align=center]
    
  \node (start)     
            [quantity]              {$e^{\frac12\! \sum u\,}\phi_{_0}$};
  \node (gofu)     [process, right of=start,yshift=1.cm,xshift=1cm]         {$g$};
  \node (E)        [process, right of=start,yshift=-1.cm,xshift=1cm]   		{$E$};
  \node (Vxc)      [quantity, right of=gofu,yshift=-2.0cm,xshift=1.4cm]   	{$\begin{array}{cc}
    \frac{\delta E[u]}{\delta u}=0  \\
    \quad\Updownarrow \\
    \frac{\delta E[g]}{\delta g}=0  \\
  \end{array} $ };
  \draw[->]            (start) -- (gofu) node[midway,above,sloped,fill=none]{$g[u]$};
  \draw[->]            (start) -- (E) node[midway,above,sloped,fill=none]{$E[u]$};
  \draw[->]     (gofu) -- (E) node[midway,right,fill=none]{$E[g]$};
  \draw[dashed,->]      (E) -- (Vxc);
  \begin{pgfonlayer}{background}
  \fill [orange!30] (-1.,-2.) rectangle (2.5,2) ;
  \fill [green!30] (2.5,-2.) rectangle (6.9,2);
  \node[fancytitle, right=10pt] at (-0.6,1.5) {FHNC/$\infty$};
  \node[fancytitle, right=10pt] at (3.2,1.3) {Euler-Lagrange \\ $\delta E=0$};
  \end{pgfonlayer}
\end{tikzpicture}
 \caption{Left: Graphical expansions provide an exact map from the JF correlations \(u_2\) to the observables $g[u_2]$ (pair distribution function) and $E[u_2]$ (energy). Right: The optimal $u_2$  minimizes \(E\), yielding the EL equation. 
 This can equivalently be interpreted as a pair density functional theory.}
 \label{fig:FHNC}
\end{figure}

\section{Theory \label{sec: Theory}}
\subsection{Exact EL equations \label{ssec: Exact}}


The functional variation (`HNC-EL') in the case of bosons gives a Schr\"odinger-like equation for \(\sqrt{g(r)}\), with effective potentials from sums of diagrams.
Pursuing the same route for fermions makes the theory much more cumbersome: exchange effects cause countless additional diagrams. 

For bosons, if elementary diagrams are neglected (`HNC/0') as their consistent treatment via two-body kernels is topologically unfeasible, two self-consistent equations arise, one being algebraic in direct, the other one in reciprocal space. 
In contrast, the corresponding fermion (`FHNC/0') result consists of eight plus eight coupled equations in real and reciprocal space.

This intricacy makes it difficult to identify FHNC-expressions with corresponding PT diagrams.
A promising route due to Krotscheck \cite{Krotscheck2000}, ensuring the correct long wavelength limit, is the `sFHNC'. Its momentum space EL-equation, \eqref{eq: EL_q}, is understood as the sum of ring diagrams (cf.\ appendices \ref{app: sFHNC},\ref{app: wI}).

Whereas that approach is rooted in approximating the static structure factor \(S(q)\) for small wave vectors \(q\),
we here derive a \textit{real space} formulation, arriving at a boson-like parquet sum \cite{Jackson1985}, where the ladders and rings are supplemented by a correction for Fermi statistics.

The Slater exchange function \(\ell(r)\!\equiv\! l(rk_{\scriptscriptstyle\mathrm F})\), i.e.\ the density matrix of a system with Fermi momentum \(k_{\scriptscriptstyle\mathrm F}\) and degeneracy factor \(\nu\) determines the non-interacting pair distribution function \(g_{\scriptscriptstyle\mathrm F}(r) = 1-
\ell(r)/\nu\), accounting for the Pauli exclusion hole.
We start with the exact FHNC expression for the pair distribution function \cite{Fantoni1975}, 
\begin{subequations}
\begin{equation}\label{eq: pairdistribution fhnc}
 g(r) =\,  [1+\Gamma_{\!\scriptscriptstyle\mathrm{dd}}(r)]\, [g_{\scriptscriptstyle\mathrm F}(r) +g_{\scriptscriptstyle\mathrm{ee}}(r)] \;,
\end{equation}
\begin{align}
 g_{\scriptscriptstyle\mathrm{ee}}=\; 
 &2\ell\,(N_{\scriptscriptstyle\mathrm{cc}\!}+E_{\scriptscriptstyle\mathrm{cc}\!})
  \,-\,\nu(N_{\scriptscriptstyle\mathrm{cc}\!}+E_{\scriptscriptstyle\mathrm{cc}\!})^2 \,+
 \nonumber\\
 &  N_{\scriptscriptstyle\mathrm{ee}\!}+E_{\scriptscriptstyle\mathrm{ee}\!} \,+\,
  2(N_{\scriptscriptstyle\mathrm{de}\!}+E_{\scriptscriptstyle\mathrm{de}\!}) \,+\,
   (N_{\scriptscriptstyle\mathrm{de}\!}+E_{\scriptscriptstyle\mathrm{de}\!})^2 \;.
\label{eq: gee exact}\end{align}
\label{eq: exact g}\end{subequations}
Here, we introduced the FHNC classification \cite{Clark1979} where the capi\-tal letters $N_{i}$, $X_{i}$, $E_{i}$ denote nodal, non-nodal and ele\-men\-tary diagrams, and \(\Gamma_{\!i}\!\equiv N_i\!+\!X_i\). The subscript \(i\in \{ \mathrm{ dd, de, ee, cc} \}\) specifies the exchange structure. Equation \eqref{eq: exact g} follows from the diagrammatic rules, which also give the relations between the various ingredients. An important example are the dd product graphs,
\begin{equation}
 \label{eq: Gu2}
  \Gamma_{\!\scriptscriptstyle\mathrm{dd}}(r)=\> \exp{\!\big[u_2(r)+N_{\scriptscriptstyle\mathrm{dd}}(r)+E_{\scriptscriptstyle\mathrm{dd}}(r)\big]}-1 \;.
\end{equation}
If all elementary graphs \(E_i\) are omitted, this route yields the first set of 8 coupled FHNC/0 equations \cite{Polls2002}. They can be solved self-consistently, if \(u_2\) in \eqref{eq: Gu2} is known.

To find the optimal \(u_2\) by energy minimization, the JF ground state energy per particle, \(e\,\),  is obtained as \cite{Krotscheck2000} 
\begin{subequations}\begin{align}
   e =\;\;& t_{\scriptscriptstyle0} + 
   \frac{\rho}{2}\!\int\! d^3r\> g(r)\, V_{\scriptscriptstyle\mathrm{JF}}(r) \,
   +\, t_{\scriptscriptstyle\mathrm{JF}} \;,
 \\
   &V_{\scriptscriptstyle\mathrm{JF}}(r) \,=\, v(r)-\frac{\hbar^2}{4m}\nabla^2 u_2(r) \;,
\end{align} 
with \(t_{\scriptscriptstyle0}\!\equiv 3\hbar^2k_{\mathrm F}^2/10m\)\,(\(m\) is a particle's mass),
and $V_{\scriptscriptstyle\mathrm{JF}}$ known as `Jastrow-Feenberg interaction'. 
The last term,
\begin{equation}
 t_{\scriptscriptstyle\mathrm{JF}} =\; -\frac{\hbar^2\rho}{8m \nu}
 \!\int\! d^3r\> \Gamma_{\!\scriptscriptstyle\mathrm{dd}}(r)\,\nabla^2 \ell^2(rk_{\scriptscriptstyle\mathrm F})
 \,+\, t_{\scriptscriptstyle\mathrm{JF}}^{(3)} \;,
\end{equation}
where \(t^{(3)}_{\scriptscriptstyle \mathrm{JF}}= 
 t^{(3\mathrm a)}_{\scriptscriptstyle \mathrm{JF}} + t^{(3\mathrm b)}_{\scriptscriptstyle \mathrm{JF}}\,\), 
 includes pair- as well as three-body exchange contributions:
\begin{align}
 t^{(3\mathrm a)}_{\scriptscriptstyle \mathrm{JF}} =\; 
 &\frac{\hbar^2\rho}{4m}\!\int\! d^3r\> \Gamma_{\!\scriptscriptstyle\mathrm{dd}}(r)\,
    \big(N_{\!\scriptscriptstyle\mathrm{cc}}(r)+E_{\!\scriptscriptstyle\mathrm{cc}}(r)\big)
 \,\nabla^2 \ell(r)
 \\
 t^{(3\mathrm b)}_{\scriptscriptstyle \mathrm{JF}} =\;
 &\frac{\hbar^2\rho^2}{8m \nu^2}\!\int\!\! d^3r_{\!_{12}}d^3r_{\!_{13}}\> 
 \Gamma_{\!\scriptscriptstyle\mathrm{dcc}}(\mathbf{r}_{\!_{1}};\mathbf{r}_{\!_{2}},\mathbf{r}_{\!_{3}}) 
 \nonumber \\
 &\hspace{2.7cm}\times\big(\nabla_{\!_1}\ell(r_{\!_{12}})\big)\,\big(\nabla_{\!_1}\ell(r_{\!_{13}})\big) \;,
\end{align}
\label{eq: ec_FHNC}
\end{subequations}
(\(\nabla_{\scriptscriptstyle1}\) denotes differentiation with respect to \(\mathbf r_{\scriptscriptstyle1}\));
$\Gamma_{\!\scriptscriptstyle\mathrm{dcc}}$ collects all three-point diagrams \cite{Krotscheck2000}, where the distinct coordinate \(\mathbf r_{\scriptscriptstyle1}\) has no exchange line but is connected with each of \(\mathbf r_{\scriptscriptstyle2}\) and \(\mathbf r_{\scriptscriptstyle3}\) via a path that does not go through the respective other one,
and a continuous exchange path exists between \(\mathbf r_{\scriptscriptstyle2}\) and \(\mathbf r_{\scriptscriptstyle3}\).

Requiring the variation of the energy with respect to $u_2$ to vanish leads to the EL equation 
\begin{align}
\label{eq:euler}
 \frac{\hbar^2}{4m}\nabla^2 g(r) \, &=
 \int\! d^3\bar r\>V_{\scriptscriptstyle\mathrm{JF}}(\bar r)
     \frac{\delta g(\bar {\mathbf r})}{\delta u_2(\mathbf r)}  \,+\, 
 \frac{2}{\rho}\frac{\delta t_{\scriptscriptstyle\mathrm{JF}}}{\delta u_2(\mathbf r)}  \nonumber \\
&\equiv g'(r) \;.
\end{align}
The rhs of the first line formally defines $g'$. It is generated diagrammatically by replacing \cite{Krotscheck2000}, in turn, 
\begin{itemize}
    \item[1.] every correlation line by $V_{\scriptscriptstyle\mathrm{JF}}(r)\,e^{u_2(r)}$  
    \item[2.] every connected pair of exchange lines by\\ \(\frac{\hbar^2}{8m}\nabla_i^2 \,\ell(r_{\!_{ij}})\ell(r_{\!_{ik}})\,\).
\end{itemize}
Applying these graphical rules, we obtain an expression for \(g'(r)\) in terms of \(\Gamma_{\!\scriptscriptstyle\mathrm{dd}}'\) and
\(g_{\scriptscriptstyle\mathrm{ee}}'\), where (as in the following) all primed quantities are constructed by employing these same rules. Inserting this definition of \(\Gamma_{\!\scriptscriptstyle\mathrm{dd}}'\) and eliminating \(u_2\) in favor of \(\,g\,\) with eqs.\,\eqref{eq: exact g}-\eqref{eq: Gu2} leads to a differential equation for \(g\) (for details, see appendix \ref{app: Detail}). 

The resulting exact EL equation, \eqref{eq: EL exact}, is Schr\"odinger-like in $r-$ space,
\begin{equation}
  \bigg[-\frac{\hbar^2}{m}\nabla^2 + v + w_{\scriptscriptstyle\mathrm I} + V_{\scriptscriptstyle\mathrm E} + V_{\!\scriptscriptstyle\mathrm{ee}} + 
  V_{\scriptscriptstyle\mathrm F}\, \bigg]_{\phantom{\big|}\!\!} \sqrt{g} \>=\, 0 \;,
\label{eq: realspace eqFHNC}\end{equation}
where 
\begin{equation}
  \qquad\qquad V_{\scriptscriptstyle\mathrm F} \equiv\,
   \frac{\hbar^2\nabla^2\sqrt{g_{\scriptscriptstyle\mathrm F}}}{m\sqrt{g_{\scriptscriptstyle\mathrm F}}} \;, \label{eq: VF def}
\end{equation}
and \(V_{\scriptscriptstyle\mathrm E}\,,\,V_{\!\scriptscriptstyle\mathrm{ee}}\) denote the contribution of dd elementary diagrams and the exchange correction, respectively (see eqs.\,\eqref{eq: eff VEr} and \eqref{eq: eff Veer}).
The `induced interaction' \(w_{\scriptscriptstyle\mathrm I}\) (derived in \eqref{eq: deriv wI VE} and further elucidated in Appendix \ref{app: wI}) is formally defined identically to that for bosons,
\begin{equation}\label{eq: wI from N}
 w_{\mathrm I}(r) \,\equiv\, \frac{\hbar^2}{4m}\nabla^2 N_{\scriptscriptstyle\mathrm{dd}}(r) + N_{\scriptscriptstyle\mathrm{dd}}'(r) \;.
\end{equation}

\subsection{Leading order ladder FHNC-EL approach \label{ssec: Approximations}}

The simplest fermionic approach is to neglect both, \(V_{\scriptscriptstyle\mathrm E}\) and \(V_{\!\scriptscriptstyle\mathrm{ee}}\) altogether, keeping  only \(V_{\scriptscriptstyle\mathrm F}(r)\),
\begin{equation}
  \bigg[-\frac{\hbar^2}{m}\nabla^2 + v(r) + w_{\scriptscriptstyle\mathrm I}(r) + 
  V_{\scriptscriptstyle\mathrm F}(r)\, \bigg]_{\phantom{\big|}\!\!} \sqrt{g(r)} \>=\, 0 \;.
\label{eq: ladplus eqFHNC}\end{equation}
All effects of the surrounding medium on a pair of particles is then contained in only two corrections to the bare \(v(r)\) in \eqref{eq: realspace eqFHNC}, namely \(w_{\scriptscriptstyle\mathrm I}(r)\) induced by correlations, and \(V_{\scriptscriptstyle\mathrm F}\) arising from the Pauli principle.

So far, we defined \(w_{\mathrm I}\) via the nodal diagrams. Attempting to derive an expression along this route, though further improving \(g(r\!\to\!0)\), does not change the long wavelength behavior. 
We therefore choose to instead incorporate the relevant \(q\!\to\!0\) terms by using Krotscheck's sFHNC expression \cite{Krotscheck2000}, 
\begin{equation}
 \tilde{w}_{\mathrm I}(q)=\, -\frac{\hbar^2 q^2}{4m}
 \left[\frac1{S(q)}-\frac1{S_{\mathrm{F}}(q)}\right]^2\left[\frac{2S(q)}{S_{\mathrm{F}}(q)}+1\right]\;,
 \label{eq: eff interactionFHNC wI}
\end{equation}
with the free static structure factor \(S_{\mathrm{F}\,}\). (As \(w_{\mathrm I}(r)\) of \eqref{eq: eff interactionFHNC wI} is finite at the origin, it ensures the cusp condition \cite{Kimball1973}.) 

Equations (\ref{eq: ladplus eqFHNC}) and (\ref{eq: eff interactionFHNC wI}) constitute a closed set of equations to be solved self-consistently, correctly reproducing the non-interacting limit.

\subsection{Relation to perturbation theory \label{ssec: relation to PT}}

Before presenting numerical results, we clarify the physics contained in \(V_{\scriptscriptstyle\mathrm F}(r)\). 
Subsuming all effective interactions in 
\(V\!\equiv v+w_{\mathrm I}\!+V_{\scriptscriptstyle\mathrm E}\!+V_{\scriptscriptstyle\mathrm {ee}}\!+V_{\scriptscriptstyle\mathrm F}\,\),
and defining \(L\!\equiv -\frac{\hbar^2}{m}\nabla^2(\sqrt{g}-\!1)\), 
the \emph{exact} EL equation \eqref{eq: realspace eqFHNC} reads in real and in Fourier space
\begin{align}
  \Big[-\frac{\hbar^2}{m}\nabla^2 + V(r) \Big] \big(\sqrt{g(r)}-\!1) \,+ V(r) = 0 \;,
  \label{eq:realspaceeqHNC bosons}
 \nonumber\\
  \widetilde L(q) =\> \widetilde V(q) \,+ \int\!\frac{d^3k}{(2\pi)^3}\,
                      \widetilde V(|{\bf q}\!-\!{\bf k}|)\,\frac{\widetilde L(k)}{t(k)} 
                      ^{\phantom{\big|}} \;.
\end{align}
This is recognized as the Bethe-Goldstone equation \cite{Lipparini2003} for bosons.
Recall that \(V_{\scriptscriptstyle\mathrm F}\) is completely independent of interactions. 
Thus, even for free fermions, the FHNC-EL can be formulated as a bosonic ladder equation with a potential originating from the Slater determinant.


The similarity is no coincidence, as the optimization of the correlation function $u_2$ happens independently of the individual states in the Fermi sea; only their sum in the density matrix $\ell$ enters the optimization\footnote{The \(q\!\to\!0\) expansion in sFHNC, where the fermionic Lindhard function is approximated by the boson-like $\chi^0_{\scriptscriptstyle\mathrm{CA}}$ \eqref{eq:chi_CA}, also shares this benefit of the state-independent Jastrow ansatz.}.

Accepting the interpretation of eq.\,\eqref{eq: realspace eqFHNC} as a ladder sum, $w_{\rm I}$ is conjectured to sum the rings (bubbles) of PT. Indeed, in appendix \ref{app: wI} we demonstrate how the sFHNC form \eqref{eq: eff interactionFHNC wI} for $\tilde w_{\rm I}$ follows from treating the rings in the 
single-pole approximation.
Both perspectives clearly corroborate the evidence that \eqref{eq: ladplus eqFHNC} combined with \eqref{eq: eff interactionFHNC wI} is an approximate self-consistent sum of ladders as well as rings, therefore denoted here as ``ladder$^+$ approximation''.

Strategies of improvement are evident: specifically, in every step of the self-consistency cycle one could replace the bosonic with the full fermionic propagator. This may answer which precise approximations in PT reproduce the EL eq.\,\eqref{eq: realspace eqFHNC}. But even the present approach, derived from an alternative formalism, gives an alternative perspective and thus can open the door for new approximation schemes in PT (in particular for ladder sums), which are hard to motivate from PT alone.

\subsection{Three bosonic FHNC-EL approaches \label{ssec: 3 bosonic FHNC-EL}}

Note that two exact relations connect the FHNC quantities $N_i$ and $X_i$ 
with the static structure; the real space eq.\,\eqref{eq: pairdistribution fhnc} and the momentum space equation, see e.g. eq. (2.7) in ref. \cite{Krotscheck2000}. 
Of course, the exact EL eq.\,\eqref{eq: realspace eqFHNC} for the optimal \(g(r)\) from a Jastrow ansatz is equivalent to its momentum space counterpart. Approximations, necessary in practice, usually break the equivalence of \({\bf r}\)- and \({\bf q}\)-space formulations.
A crucial point here is that (the real space) eq.\,\eqref{eq: pairdistribution fhnc} with \(g_{\scriptscriptstyle\mathrm{ee}}\!=\!0\) better approximates the ladders, whereas eq.\,\eqref{eq:Sq} (in momentum space) is superior for the bubbles. We here test the performance the following three simple, boson-like types of FHNC/0-EL approaches:
\begin{itemize}\setlength{\itemindent}{0.8cm}
 \item[ladder$^{\scriptscriptstyle+}$:]
   self-consistent solution of \eqref{eq: ladplus eqFHNC} and \eqref{eq: eff interactionFHNC wI}
 \item[s{\small FHNC}:]
   self-consistent solution of \eqref{eq:Sq}-\eqref{eq:Vph from wI} and \eqref{eq: eff interactionFHNC wI} 
 \item[b{\small FHNC}:]
   Kallio's \cite{Kallio1996} suggestion (see below) 
\end{itemize}
For remarks on the implementation, see appendix \ref{app: Implementation}.

On purpose, and in hindsight of the theory's extension to periodic structures, we refrain from any sophisticated modifications, but test the most `naked' versions of ladder$^+$ and sFHNC.
Note, however, that the Fourier transform of a good approximation for \(S(q)\) from certain FHNC equations may give \cite{Krotscheck77} an inadequate \(g(r)\). We here stress that applications of sFHNC can be done in a refined manner (cf.\ end of appendix \ref{app: sFHNC}). Our intention is to figure out, where the three approaches work best, so that they can complement each other.

The version introduced by \citeauthor{Kallio1996} \cite{Kallio1996} enforces the non-interacting fermion limit on the bosonic HNC-EL and was further motivated from a density functional theory (DFT) perspective \cite{DAPT03_analytic}. For brevity, we call it bFHNC; (its implementation is analogous to that of ladder$^+$, cf.\ appendix \ref{app: Implementation}). This variant has been sucessfully applied to various charged systems before. 

\section{Results}

We now test the above approaches for the HEG. The basic assessment of a theory's accuracy is to compare the correlation energy per particle, \(e_{\mathrm c}(r_s)\), with simulation results. We obtain it by coupling constant integration from the pair distribution function $g_{\bar r_s}(r)$  at a density parameter $\bar r_s$
\begin{equation}\label{eq:ec}
 e_{\mathrm c}(r_s)=\frac{3}{8\pi \, r_s^2}
 \int\limits_0^{r_s}\!\frac{d\bar{r}_s}{\bar{r}_s^2} 
  \int\! \frac{d^3r}{a_0^3} \; v(r)\,\big[g_{\bar{r}_s}(r)-g_{\scriptscriptstyle\mathrm F}(r)\big] \;.
\end{equation}
This procedure has the benefit that all three different approximations are compared on the same level\footnote{The sFHNC energy can also be directly calculated from eq.\,\eqref{eq: ec_FHNC}; the bHNC functional is unknown. 
The ladder$^+$ combination of eqs.\,\eqref{eq: ladplus eqFHNC}-\eqref{eq: eff interactionFHNC wI} cannot be uniquely mapped on expression \eqref{eq: ec_FHNC} either.}, only based on their pair distribution functions.

\begin{figure}[t]
  \includegraphics[width=8.6cm,keepaspectratio=true]{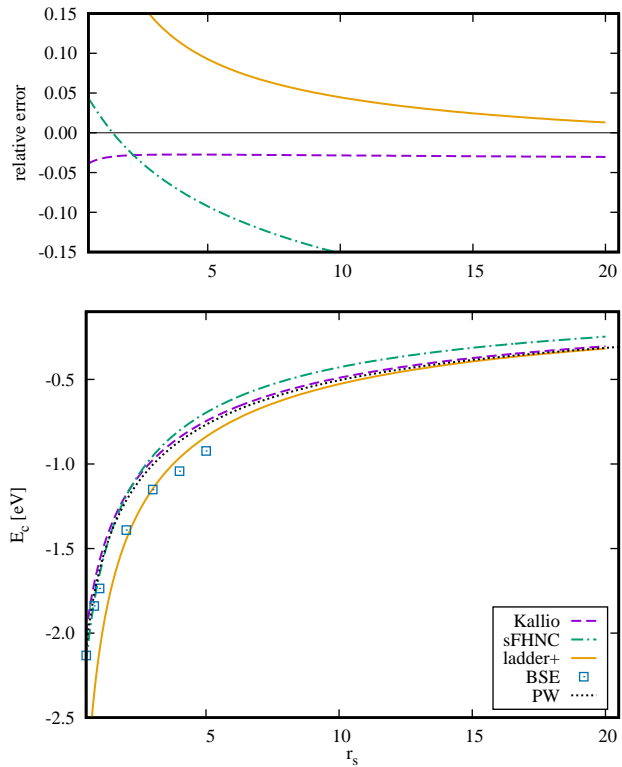}
  \caption{Correlation energy per particle versus density parameter for three FHNC-EL variants. Bottom: bFHNC (dashed violet line), sFHNC ring-summation (dot-dashed cyan line) and ladder$^+$ approach (full orange line), compared to BSE \cite{Maggio2016} (boxes) 
  and parametrized MC data \cite{Perdew1992,Ceperley1980} (dotted black line). 
  Top: Relative error with respect to MC.
  \label{fig:Ec}}
 \end{figure}

The correlation energies of the ladder$^+$, the sFHNC, and Kallio's bFHNC are compared to MC results in fig.\,\ref{fig:Ec}. The bFHNC performs best, with an error of a few percent over the whole $r_s$ range.
The sFHNC works reasonably well for metallic densities (how the deviation for \(r_s\!\to\!0\) can be corrected is reported in appendix \ref{app: sFHNC}). 
In accordance with the PT knowledge that ladder summations are crucial in the highly correlated regime \cite{Lipparini2003}, the ladder$^+$ curve is seen to become superior for \(r_s\!\gtrsim15\). 
  
We also compare with the seminal BSE results of Maggio and Kresse \cite{Maggio2016}. 
They evaluated the four-point particle-hole ladders with a static RPA screened interaction. 
It appears promising to replace this by FHNC (CBF) interactions for the particle-hole ladder \cite{Panholzer2009,Panholzer2010}. 


Despite the substantial simplifications of the various propagators in all three FHNC/0-EL approaches, they perform notably well. This indicates that their effective interactions are of high quality, accounting for the most relevant physics in the respective density regime.

\begin{figure}[t]
  \includegraphics[width=8.6cm,keepaspectratio=true]{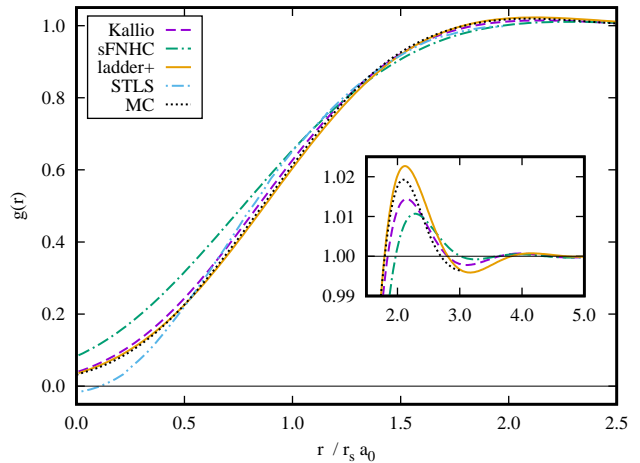}
  \caption{Pair distribution function of the paramagnetic HEG at $r_s\!=\!5$ in the ladder$^+$, bFHNC and sFHNC aproaches (line types as in fig.\,\ref{fig:Ec}), 
  compared to a fit of MC data (Spink et al.\,\cite{Spink2013}, dotted black curve);
  STLS data taken from \cite{Singwi1969} (double-dot-dashed blue line). Inset: first maximum. 
  }
 \label{fig:gr}
 \end{figure}

An accurate energy does not guarantee a high quality wave function or static structure. As is well known, the STLS \cite{Singwi1969} yields excellent correlation energies, but fails for \(g(r\!\to\!0)\), yielding negative values for $r_s\!\gtrsim\!4$. The $r_s\!=\!5$ pair distribution functions are displayed in fig.\,\ref{fig:gr}; we compare with the most recent simulations of Spink et al.\,\cite{Spink2013}.
Remarkably, the ladder$^+$ approximation is closer to MC than the bFHNC, which is of high quality, too. 
The sFHNC and STLS clearly deviate from the MC data; the inset shows that their nearest-neighbour peak position \(r_{\rm m1}\) is dissatisfying. 

Per construction, ladder$^+$ performs insufficiently for \(r_s\!\to\!0\), this error is carried over and accumulated by the coupling constant integration (\ref{eq:ec}), explaining the deviations of \(e_{\mathrm c}\) from MC.
We conclude that $g(r)$ is most satisfying in ladder$^+$, but the kinetic energy is less accurate. 

\begin{figure}[t]
  \includegraphics[width=8.6cm,keepaspectratio=true]{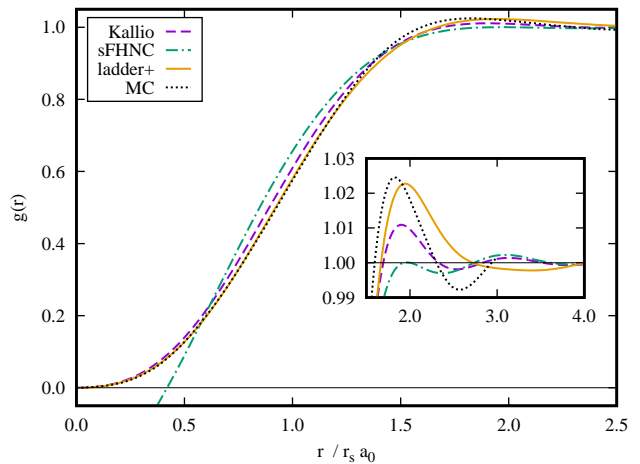}
  \caption{Same as fig.\,\ref{fig:gr}, but for the ferromagnetic HEG. The STLS curve is omitted, lying    significantly below the sFHNC.}
  \label{fig:gr_pol}
 \end{figure}

The fully spin polarized HEG is depicted in fig.\,\ref{fig:gr_pol}, again for $r_s\!=\!5$. For small $r$ the spin polarized ladder$^+$ \(g(r)\) is on top of the MC benchmark curve \cite{Spink2013}. 
The behavior of the first peak position \(r_{\rm m1}\), foreshadowing the Wigner crystal's nearest neighbours, is interesting.
The MC peak has the lowest \(r_{\rm m1}\); in the paramagnetic HEG the ladder$^+$ value agrees closely, its ferromagnetic first maximum is too far right. 

Note that these discrepancies are only \(\sim\)2\%. For the two-dimensional (2D) HEG, where correlation effects are more pronounced, neglecting elementary diagrams and triplet correlations leads to larger deviations (see the review by \citeauthor{Asga08_MBeffects} \cite{Asga08_MBeffects}). This trend continues to 1D, but still the ladder$^+$ serves as a reasonable starting point \cite{Panholzer2017}. Similar expectations hold for 
partial spin-polarization \cite{kreil2018resonant}.

\begin{figure}[t]
  \includegraphics[width=8.6cm,keepaspectratio=true]{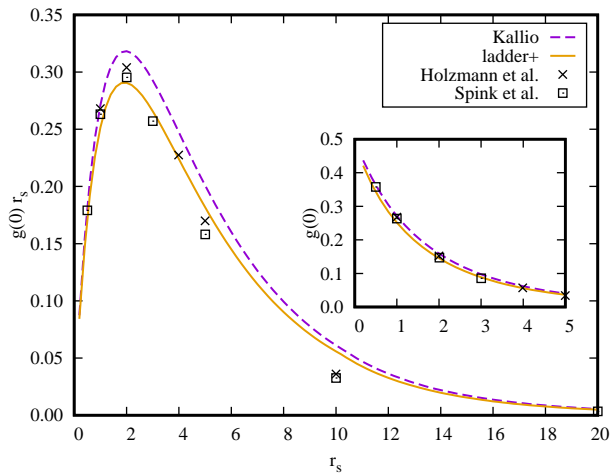}
  \caption{On-top pair distribution function $g(0)$ multiplied by $r_s$ for the paramagnetic HEG. The Kallio and ladder$^+$ approach is compared to the MC results of Spink et al.\cite{Spink2013} and Holzmann et al.\,cite{Holzmann2011}. The inset demonstrates the correct behavior for $g(0)$ for small $r_s$. 
  }
  \label{fig:gontop}
 \end{figure}

\begin{figure}[t]
  \includegraphics[width=8.6cm,keepaspectratio=true]{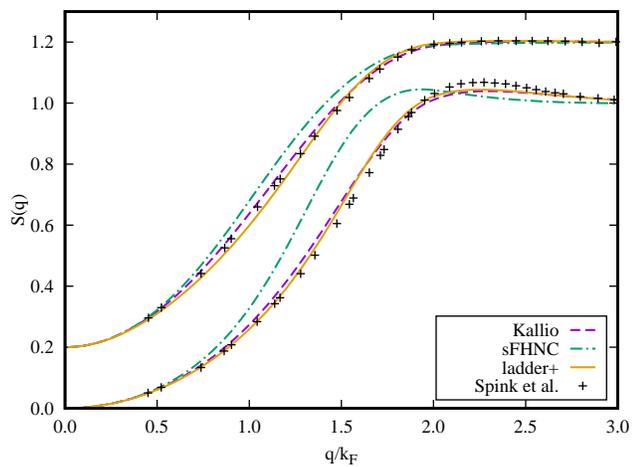}
  \caption{The static structure function of the three FHNC approximations at $r_s=20$ and $r_s=5$ (shifted up by 0.2), compared to the respective MC result \cite{Spink2013}.}
  \label{fig:Sq}
 \end{figure}

In fig.\,\ref{fig:gontop} we compare the on-top pair distribution function \(g(0)\). The overall agreement of both, the ladder$^+$ and the bFHNC values with the MC results is good, with a slight superiority of the former. 
The static structure factor in fig.\,\ref{fig:Sq} demonstrates for all approximations the correct $q^2$ behavior for long wavelengths, confirming the overall picture already discussed.

Of course, some of the less satisfactory features of the above approaches can be removed: 
Takada \cite{YosT09:newgeneral} suggested an improved STLS scheme, in sFHNC the small $r$ of $g(r)$ behavior can be corrected as indicated at the end of appendix \ref{app: sFHNC}. 
Here, we want to compare them in their `most naked' form and it turns out that even these FHNC versions perform highly satisfactory.

\begin{figure}[t]
  \includegraphics[width=8.6cm,keepaspectratio=true]{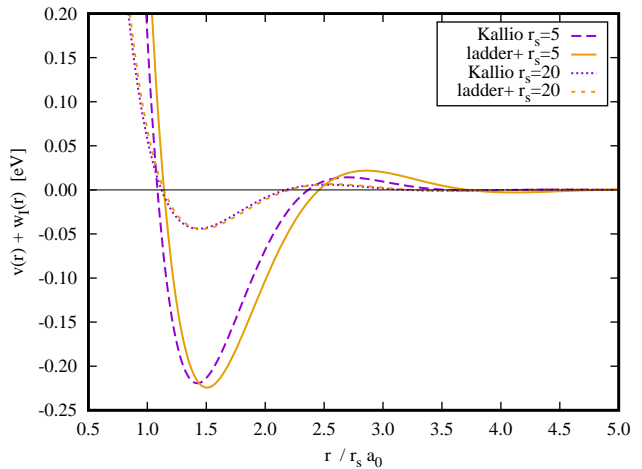}
  \caption{Effective interaction $v(r)+w_{\rm I}(r)$ for the optimized correlations at $r_s\!=\!5$ and $r_s\!=\!20$. The corresponding bFHNC potential is  $\,v + w_{\scriptscriptstyle\mathrm{IB}}-w_{\scriptscriptstyle\mathrm{IBF}}$ (see appendix \ref{app: Implementation}).}
  \label{fig:wI}
 \end{figure}

Finally, in fig.\,\ref{fig:wI} we show the bare plus induced interactions, $v(r)+\wI(r)$, as it appears as driving term in eq. \eqref{eq: realspace eqFHNC}. Since bFHNC has an EL eq. of similar form, we compare to the corresponding expression  
$\,v + w_{\scriptscriptstyle\mathrm{IB}}-w_{\scriptscriptstyle\mathrm{IBF}}$ (see appendix \ref{app: Implementation}). 
%
Most prominent is the minimum at 1.5$\,r_s a_0$, again a precursor of the Wigner crystal. Although in absolute units the minimum gets
smaller with higher $r_s$, it has to be compared with the kinetic energy which scales with $r_s^{-2}$. Thus the effective depth becomes larger, correctly leading to a higher nearest neighbour peak in \(g(r)\).


\section{Systematic improvements}

Having demonstrated that the ladder$^+$ yields generally good, and for strong correlations excellent
results, we here show how it can be \emph{systematically} improved --- another strength of the present theory.

\subsection{Jastrow--Feenberg contributionss}

\textit{Additional diagrams:} Most obvious is to include $V_{ee}^0$, the superscript 0 indicating that elementary graphs in $V_{\scriptscriptstyle\mathrm{ee}}$ are neglected\footnote{Consequently, $\wI\,$, obtained from the FHNC equations for $N$ and $N'$, then takes a more complicated form than eq.\,\eqref{eq: eff interactionFHNC wI}}.
For the non-bosonized FHNC this was done by Lantto \cite{Lantto1980}, with a similar result as our ladder$^+$ approximation. This further justifies the simplified treatments.
Including elementary diagrams is then the next step.

\textit{Higher order correlations:}
Using not only pair correlations in the wave function $\psi=F \phi_{\scriptscriptstyle0\,}$, but adding triplet correlations $u_3({\bf r}_1,{\bf r}_2,{\bf r}_3)$ (and possibly higher order $u_n$), further amends the theory. 
The inclusion of both, the diagram \(E_4\) and triplets \(u_3\), significantly improves the bFHNC results for the 2D HEG \cite{Asga08_MBeffects}.

\subsection{Correlated basis functions}

While for bosons by adding $u_n({\bf r}_1,\dots,{\bf r}_n)$ of arbitrarily high order the exact ground state is reached, the unknown nodes of $\psi$ prevent this for fermions. 
The CBF framework \cite{BKro02} remedies this problem. 
Particle-hole excitations of the Slater determinant, $\phi_{ph}\!=a^{\dagger}_p a_h\,\phi_0\,$, correlated by $F$ of the FHNC ground state, form a correlated basis, allowing to obtain the true ground state.
Compared to standard PT, the CBF convergence is much faster (but derivations become more involved). We exemplify the principle of the method for the ring diagrams. Their PT summation yields the RPA. The analogous route in CBF gives a similar density response function,
\begin{equation}\label{eq: cRPA}
 \chi_{\mathrm{c\scriptscriptstyle RPA}}(q,\omega)\,=\,
 \frac{\chi_0(q,\omega)}{1-\widetilde V_{\!\scriptscriptstyle\mathrm{ph}}(q)\,\chi_0(q,\omega)} \;,
\end{equation}
with the bare \(\tilde v(q)\) replaced by the particle-hole irreducible interaction $\widetilde V_{\!\scriptscriptstyle\mathrm{ph}}=\widetilde X_{\scriptscriptstyle\mathrm{dd}}'-\widetilde X_{\scriptscriptstyle\mathrm{dd}\,}\hbar^2q^2/4m$. (The derivation \cite{Panholzer2010} is a bit tedious; 
an explicit expression for $\widetilde V_{\scriptscriptstyle\mathrm{ph}}$ in sFHNC  is given in eq.\,\eqref{eq:Vph from wI}).

How the induced \(w_{\mathrm I}\) of sFHNC follows from the plasmon pole approximation\footnote{Using the full Lindhard function does not give an explicit \(\widetilde V_{\!\scriptscriptstyle\mathrm{ph}}(q)\,\).} of \(\chi(q,\omega)\) together with the \(m_0\) sum rule is outlined in appendix \ref{app: wI}. There, we also delineate the close relation of \(\widetilde V_{\scriptscriptstyle\mathrm{ph}}\) with the ladders.

The cRPA response function \eqref{eq: cRPA} significantly improves the bare RPA dynamics. It should be compared to time-dependent DFT (TDDFT). An explicit non-local approximation for the exchange-correlation kernel is obtained via $\tilde f_{\rm xc}\approx \widetilde V_{\scriptscriptstyle\mathrm{lad}}=\widetilde V_{\!\scriptscriptstyle\mathrm{ph}\!}-\tilde v$, to be tested against other static kernels \cite{Olsen2013,Corradini1998}. 

Similarly, we propose an upgraded treatment of the fermionic Bethe-Goldstone equation in PT. 
Since $\wI$ contains, though approximately, a large class of diagrams, using the interaction $v+w_{\mathrm I}$ as a driving term, the corresponding ladder sum is expected to be superior to that with the bare $v$, and to give quicker convergence if used as a starting point in more refined summations.
The latter would allow to further assess the accuracy of the above boson-like FHNC schemes (ideally to be compared to a fully self-consistent sum of fermion rings and ladders, still a numerically challenging task). 

Other classes of PT diagrams can be approximated by FHNC summations, too. Self-energy diagrams, contained in $g_{ee}$, are neglected in the present treatment. Again, a combination of FHNC and PT gives good results \cite{Seydi2018}.

\subsection{Extension to inhomogeneous systems}
The low computational demands of the method suggests a generalization to inhomogeneous systems. Numerically, the huge step from a one dimensional $g(r)$ to a six dimensional $g(\mathbf{r}, \mathbf{r}')$ makes exploiting symmetry unavoidable. As demonstrated in Ref.\,\cite{Panholzer2017}, the solution of the FHNC equations for periodic systems is numerically feasible and further work in that direction is in progress. 

Our present study adds value and reliability to that attempt. First, the relation provided to PT is particularly useful for describing excited state properties of realistic systems.
Second, the non-local behaviour of two-point quantities is kept, i.e., neither local nor semi-local approximations are needed --- in contrast to the local density approximation (LDA) or the generalized gradient approaches in DFT. 
Starting from TDDFT and utilizing the adiabatic connection formula partly removes 
this locality with promising results \cite{Olsen2013,Patrick2015}; however, the non-locality of $f_{\rm xc}$ still has to be approximated. The  inhomogeneous version of the theory yields a non local $f_{\rm xc}$ at the same level of approximation as the homogeneous version.
Third, it allows to account for or estimate the different approximations done, by calculating, e.g., $V_{\scriptscriptstyle\mathrm{ee},0}$, low order contributions to $V_E$ or CBF corrections.

\section{Conclusion}

We here demonstrated the strength of uncomplicated FHNC-EL versions. Their key advantage is to be based on \textit{functional} optimization, thus yielding a parameter-free, unbiased result for the ground state structure. The intricacy of the full fermion version is avoided by neglecting elementary diagrams and exchange corrections $g_{\scriptscriptstyle\mathrm {ee}}$. 
The resulting self-consistency equations exhibit physical transparency and share the low computational demand with classical HNC. From a practical perspective, they are thus extremely efficient, while nevertheless yielding rather accurate results.

The HEG served for testing and comparing the following specific approaches: the bFHNC formulated by Kallio and Piilo \cite{Kallio1996}, the sFHNC introduced by Krotscheck \cite{Krotscheck2000}, and 
an FHNC-EL version that we newly developed for the short range region and the low density limit of HEGs. It self-consistently sums approximated ladder and ring diagrams with emphasis on the former, motivating the term ladder$^+$ approach. 

The bFHNC method performs best in a wide density range, yielding both, pair distribution function and correlation energy close to the MC benchmark data. For high and low densities, \(g(r)\) is more accurate in sFHNC and ladder$^+$, respectively. Their additional advantage is to allow a connection with PT.
Consequently, their extension to periodic systems \cite{Panholzer2017} holds a high potential for an implementation in combination with PT algorithms for solid state physics, e.g.\ in BSE \cite{Maggio2016} .

Apart from the good performance combined with low numerical cost, we stress the following: Although derivations in the FHNC-EL formalism are intricate, they \textit{justify} the resulting equations and energy functionals. Once established, these can then be applied in a pair DFT \cite{Krot94:consistency, DAPT03_analytic, Furc04:towards, Higuchi2013}, the pair analog of conventional DFT. 


 
Finally, the possibility of coherently refining the method via CBF and generalizing the static ground state correlations to dynamic fluctuations \cite{Bohm2010, Panholzer2018}   underpins the value and utility of the FHNC approach.
\\
 
The research was supported by the Austrian science fund FWF under Project No. J 3855-N27.\\

\appendix

\section{sFHNC \label{app: sFHNC}}

We provide a brief summary of the sFHNC as formulated by Krotscheck \cite{Krotscheck2000}.
The static structure factor $S$ and its non-interacting counterpart $\SF$ are related to the sum of all direct--direct cluster diagrams  $\widetilde\Gamma_{\!\scriptscriptstyle\mathrm{dd}}$
via
\begin{equation}\label{eq:Sq}
  S(q) = \>\SF(q)\,\big(1+ \SF(q)\,\widetilde\Gamma_{\!\scriptscriptstyle\mathrm{dd}}(q)\big) \;,
\end{equation}
which is exact for \(q\!\to\!0\). Denoting the energy of a free single particle as \(t(q)\equiv\hbar^2q^2/2m\),
the EL equation resulting from minimizing the ground state energy reads 
\begin{equation}\label{eq: EL_q}
 S(q) \,=\frac{\SF(q)}{\sqrt{1+\frac{2\SF^2(q)}{t(q)} \Vqh^{\scriptscriptstyle\mathrm{sFHNC}}(q) }} \;,
\end{equation} 
with the particle--hole irreducible interaction \cite{Krotscheck2000} 
\begin{align}
 \label{eq:Vph from wI}
  \Vph^{\scriptscriptstyle\mathrm{sFHNC}}(r)=&\,\big(1+\Gamma_{\!\scriptscriptstyle\mathrm{dd}}(r)\big)\,v(r) + 
  \textstyle\frac{\hbar^2}{m}\displaystyle\Big|\nabla\sqrt{1\!+\!\Gamma_{\!\scriptscriptstyle\mathrm{dd}}(r)}\Big|^2 \nonumber\\& +
  \Gamma_{\!\scriptscriptstyle\mathrm{dd}}(r)\,w_{\mathrm I}(r) \;.
\end{align}
Here, \(w_{\mathrm I}\) is related to the summed nodal diagrams exactly as in \eqref{eq: wI from N}.
Consistent with the approximations leading to \eqref{eq:Sq}, \(\tilde w_{\mathrm{I}\!}\) can be expressed in terms of the static structure function, resulting in eq.\,(\ref{eq: eff interactionFHNC wI}).
From \(\Gamma_{\!\scriptscriptstyle\mathrm{dd}}(r)\) the potential \(\Vph^{\scriptscriptstyle\mathrm{sFHNC}}\) is obtained with \eqref{eq:Vph from wI}, yielding \(S(q)\) via \eqref{eq: EL_q}, and then \(\!\widetilde\Gamma_{\!\scriptscriptstyle\mathrm{dd}}\) for the next iteration step from \eqref{eq:Sq}.

We emphasize, that also the sFHNC contains both, ring- and ladder diagrams in $\Vph^{\scriptscriptstyle\mathrm{sFHNC}}$, eq.\,\eqref{eq:Vph from wI}, and that these are summed in an approximate but \textit{consistent} way.

As Krotscheck \cite{Krotscheck77} 
pointed out, the Fourier transform (FT) of \(S(q)\) does not yield a physically meaningful \(g(r)\). Instead, \(g\) should be calculated differently, or additional FHNC integral equations should be included for the exchange diagrams. The importance of the latter, alternatively, can be estimated by comparing a good \(g(r)\) with the FT of \(S(q)\).

Since eq.\,\eqref{eq:Sq} is designed to exactly reproduce the FHNC $q\!\to\!0$ limit, the small $r$ regime may well lack quality.
Specifically, $g(r\!\to\!0)$ is not well approximated and violates the cusp condition. 

An ad hoc recipe \cite{Egger2011} is to obtain  $g\,$ from eq.\,\eqref{eq: pairdistribution fhnc} with $g_{\rm ee}\!=\!0$, i.e., $g\!=(1\!+\Gamma_{\!\scriptscriptstyle\rm dd})\,g_{\scriptscriptstyle\rm F}$, as it holds in ladder$^+$, but with $\Gamma_{\!\scriptscriptstyle\rm dd}(r)$ being the FT of $\widetilde\Gamma_{\!\scriptscriptstyle\mathrm{dd}}=(S-S_{\scriptscriptstyle\rm F})S_{\scriptscriptstyle\rm F}^{-2}$. We do not use this for the following reasons: First, it violates the most fundamental relation between $S(q)$ and $g(r)$, eq.\,\eqref{eq: FT Stog}. Second, though improving $g(r\!\to\!0)$, the result gets much worse for intermediate distances, $r\gtrsim 0.5\,r_sa_0$, in the present case of Coulomb systems. Third, this additional assumption spoils the simplicity and elegance of the original approach.


\section{Detailed derivation of the Euler Lagrange equation \label{app: Detail}}
 
We here provide all contributions to the exact FHNC-EL theory.
 Similar to Ref.\,\cite{Lantto1977} we stay in real space, but in contrast to their usage of Lagrange multipliers in the optimization procedure, we employ the diagrammatic rules.
 
 To get familiar with the formalism, we demonstrate the explicit procedure for obtaining the `primed' $\Gamma_{\!\scriptscriptstyle\mathrm{dd}}'$ by applying the graphic rules defined in Sec.\,\ref{sec: Theory} to the `unprimed' $\Gamma_{\!\scriptscriptstyle\mathrm{dd}}$ in eq.\eqref{eq: Gu2}. First, we replace \(\,e^{u_2}\!-\!1\,\) by $V_{\scriptscriptstyle\mathrm{JF}}\,e^{u_2}$. 
 Then we need to utilize both rules for the collection of $N_{\scriptscriptstyle\mathrm{dd}}$ and $E_{\scriptscriptstyle\mathrm{dd}}$ graphs, 
 thus defining \(N_{\scriptscriptstyle\mathrm{dd}}'\) and \(E_{\scriptscriptstyle\mathrm{dd}}'\) as sums of specific diagrams. 
 Rewriting 
 \begin{align}
     \Gamma_{\!\scriptscriptstyle\mathrm{dd}} 
     =&\left( e^{u_2}-1\right)\left(1+ N_{\scriptscriptstyle\mathrm{dd}}+\textstyle\frac12\displaystyle N_{\scriptscriptstyle\mathrm{dd}}^2+ \dots \right)e^{E_{\scriptscriptstyle\mathrm{dd}}} \\\nonumber
     &\qquad\quad+\left(1+ N_{\scriptscriptstyle\mathrm{dd}}+\textstyle\frac12\displaystyle N_{\scriptscriptstyle\mathrm{dd}}^2+ \dots \right)e^{E_{\scriptscriptstyle\mathrm{dd}}}-1 \;,
 \end{align}
 it is straightforward to apply rule 1 directly, and both rules to all sub-diagrams\footnote{The rules are used in turn for each sub-diagram; e.g. applied to $N_{\scriptscriptstyle\mathrm{dd}}^2/2$, they give $N^{\phantom{2}}_{\scriptscriptstyle\mathrm{dd}} N_{\scriptscriptstyle\mathrm{dd}}'$.}.
 Explicit expressions are obtained by invoking the rules in the FHNC equation for \(N_{\scriptscriptstyle\mathrm{dd}}\) and the chosen approximation to \(E_{\scriptscriptstyle\mathrm{dd}}\). 
 
 To keep the derivation general, we continue with the exact expression.
 Collecting all product graphs leads to 
 \begin{equation}\label{eq: Gu2primed}
     \Gamma_{\!\scriptscriptstyle\mathrm{dd}}'=(1\!+\Gamma_{\!\scriptscriptstyle\mathrm{dd}})\,
     \left(V_{\scriptscriptstyle\mathrm{JF}}+ N_{\scriptscriptstyle\mathrm{dd}}'+    E_{\scriptscriptstyle\mathrm{dd}}'  \right) \;.
 \end{equation}
This is applied analogously to all FHNC equations. The underlying definition being based on a functional derivative, this mostly amounts to taking the `ordinary' derivative of each equation, where all new primed ingredients are well-defined diagrammatic sums. 
When $g_{\scriptscriptstyle\mathrm{ee}}'$ is obtained graphi\-cally, the execution of rule 2 for two connected cc diagrams requires some care (in the literature often avoided by neglecting $t_{\scriptscriptstyle\rm JF}^{\scriptscriptstyle\rm(3b)}$, argued to be negligibly small). Some pertinent details
are discussed at the end of this section.
 
 For \(g'\) obtained from \eqref{eq: pairdistribution fhnc},
 \begin{equation}\label{eq: gprimed}
 g' =\> \Gamma_{\!\scriptscriptstyle\mathrm{dd}}'\,
  \big(g_{\scriptscriptstyle\mathrm F}\! + g_{\scriptscriptstyle\mathrm{ee}}\big)
  \,+\, \big(1\!+ \Gamma_{\!\scriptscriptstyle\mathrm{dd}}\big)\,
  \Big(\textstyle\frac{\hbar^2}{4m}\displaystyle\nabla^2 g_{\scriptscriptstyle\mathrm{F}}+g_{\scriptscriptstyle\mathrm{ee}}'\Big)
  \;,
 \end{equation}
 we can now rewrite the first term with \eqref{eq: Gu2primed} as
 \begin{equation}\label{eq: Gammapr1}
  \Gamma_{\!\scriptscriptstyle\mathrm{dd}}'\,
  (g_{\scriptscriptstyle\mathrm F} + g_{\scriptscriptstyle\mathrm{ee}})
  \>=\> g\, (V_{\scriptscriptstyle\mathrm{JF}}+ N_{\scriptscriptstyle\mathrm{dd}}' + E_{\scriptscriptstyle\mathrm{dd}}'\big) \;.
 \end{equation}
Here,  we replace $u_2$ in $V_{\scriptscriptstyle\mathrm{JF}}$ with eq.\,\eqref{eq: exact g} by
\begin{align}
 u_2
 &=\ln{\!\left(\frac{g}{g_{\scriptscriptstyle\mathrm F}+g_{\scriptscriptstyle\mathrm{ee}}}\right)} -N_{\scriptscriptstyle\mathrm{dd}}-E_{\scriptscriptstyle\mathrm{dd}} \;,
\end{align}
and obtain for the rhs of eq.\eqref{eq: Gammapr1}
\begin{align}\label{eq: deriv wI VE}
&g \bigg(v-\frac{\hbar^2}{4m}\nabla^2\Big[ 
 \ln{\!\Big(\frac{g}{g_{\scriptscriptstyle\mathrm F}\!+g_{\scriptscriptstyle\mathrm{ee}}}\Big)}
  -N_{\scriptscriptstyle\mathrm{dd}}-E_{\scriptscriptstyle\mathrm{dd}}  \Big]+ N_{\scriptscriptstyle\mathrm{dd}}' + E_{\scriptscriptstyle\mathrm{dd}}' 
  \bigg)_{\phantom{\big|}\!\!}
  \nonumber\\
  &\equiv\> g\bigg(v
  -\frac{\hbar^2}{4m}\nabla^2
  \ln{\!\Big(\frac{g}{g_{\scriptscriptstyle\mathrm F}\!+g_{\scriptscriptstyle\mathrm{ee}}}\Big)}
  + w_{\scriptscriptstyle\mathrm I}+ V_{\scriptscriptstyle\mathrm E} \bigg)
\end{align}
This proves eq.\,\eqref{eq: wI from N} for the induced potential \(w_{\scriptscriptstyle\mathrm I}\) and gives an analogous expression for another effective interaction,
\begin{equation}
  V_{\scriptscriptstyle\mathrm E} =\frac{\hbar^2}{4m}\nabla^2 E_{\scriptscriptstyle\mathrm{dd}}+E_{\scriptscriptstyle\mathrm{dd}}' \;,
  \label{eq: eff VEr}
\end{equation}due to dd elementary diagrams.
Though coinciding with the bosonic formulae, \(w_{\scriptscriptstyle\mathrm I}\) and \(V_{\scriptscriptstyle\mathrm E}\) contain many more diagrams arising from the various internal exchange lines.

Using the identity
\begin{equation}
    \nabla^2 g + g\,\nabla^2\ln{g} \,=\, 4\sqrt{g}\> \nabla^2\!\sqrt{g} \;,
\label{eq: identity}\end{equation}
the exact EL equation \eqref{eq:euler} takes the form
\begin{align}
 0 =\>& \Big(\frac{\hbar^2}{4m}\nabla^2 g -g'\Big)/\sqrt{g} \\\nonumber
 =\>&
 \frac{\hbar^2}{m}\nabla^2\!\sqrt{g} -
 \sqrt{g}\,\Bigg(\,v + w_{\scriptscriptstyle\mathrm I}+ V_{\scriptscriptstyle\mathrm E}
  +\frac{\hbar^2}{4m}\nabla^2
  \ln{\!\big(g_{\scriptscriptstyle\mathrm F}\!+g_{\scriptscriptstyle\mathrm{ee}}\big)}
  \\\nonumber
  &\hspace{3.0cm}\frac{1}{g_{\scriptscriptstyle\mathrm{F}}+g_{\scriptscriptstyle\mathrm{ee}}} \Big(\frac{\hbar^2}{4m}\nabla^2 g_{\scriptscriptstyle\mathrm{F}}+g_{\scriptscriptstyle\mathrm{ee}}'\Big)
   \bigg) \;,
\end{align}
where \(\Gamma_{\!\scriptscriptstyle\mathrm{dd}}\) on the rhs of \eqref{eq: gprimed} has been expressed with \(g_{\scriptscriptstyle\mathrm{ee}}\).

We intentionally separate the exchange contributions into the \emph{purely statistical} \(g_{\scriptscriptstyle\mathrm{F}}\) and the \emph{correlation dominated} \(V_{\scriptscriptstyle\mathrm{ee}}\,\),
 \begin{align}
 &\ln{\!\big(g_{\scriptscriptstyle\mathrm F}\!+g_{\scriptscriptstyle\mathrm{ee}}\big)} 
 = \ln{g_{\scriptscriptstyle\mathrm F}} + 
   \ln{\!\big(1+ g_{\scriptscriptstyle\mathrm{ee}}/g_{\scriptscriptstyle\mathrm F}\big)}
 \\\nonumber 
 &\frac{1}{g_{\scriptscriptstyle\mathrm{F}}+g_{\scriptscriptstyle\mathrm{ee}}} 
 \>= \frac1{g_{\scriptscriptstyle\mathrm{F}}} +\left(\frac1{g_{\scriptscriptstyle\mathrm{F}}+g_{\scriptscriptstyle\mathrm{ee}}}-\frac1{g_{\scriptscriptstyle\mathrm{F}}} \right) \;.
 \end{align}
 Collecting all terms and applying \eqref{eq: identity} to \(g_{\scriptscriptstyle\mathrm F}\) leads to
 \begin{equation}\label{eq: EL exact}
  \frac{\hbar^2}{m}\nabla^2\sqrt{g} \>=\> \left(
  v+ w_{\scriptscriptstyle\mathrm I} + V_{\scriptscriptstyle\mathrm E} + V_{\scriptscriptstyle\mathrm{ee}} + 
  \frac{\hbar^2}{m} \frac{\nabla^2 \sqrt{g_{\scriptscriptstyle\mathrm F}}}{\sqrt{g_{\scriptscriptstyle\mathrm F}}}
   \right)\sqrt{g} \;,
 \end{equation}
with the exchange correction
\begin{widetext}
 \begin{equation}
  V_{\scriptscriptstyle\mathrm{ee}} \,=\, 
  \frac{\hbar^2}{4m}\nabla^2 \ln{\!\Big(1+\frac{g_{\scriptscriptstyle\mathrm{ee}}}{g_{\scriptscriptstyle\mathrm F}}\Big)}
  \,+\, \frac{g'_{\scriptscriptstyle\mathrm{ee}}/g_{\scriptscriptstyle\mathrm F}}{1+g_{\scriptscriptstyle\mathrm{ee}}/g_{\scriptscriptstyle\mathrm F}} 
  \>-\> \frac{g_{\scriptscriptstyle\mathrm{ee}}/g_{\scriptscriptstyle\mathrm F}}{1+g_{\scriptscriptstyle\mathrm{ee}}/g_{\scriptscriptstyle\mathrm F}}  \left(\frac{\hbar^2}{4m}\frac{\nabla^2 g_{\scriptscriptstyle\mathrm F}}{g_{\scriptscriptstyle\mathrm F}}
  \right) \;.
  \label{eq: eff Veer}
 \end{equation}
\end{widetext}
The $\sqrt{g_{\scriptscriptstyle\mathrm{F}}}\,$-term in \eqref{eq: EL exact} can be viewed as a simple Fermi correction to the bosonic ladder (propagator) equation, in analogy to sFHNC, which is an approximation to PT ring diagrams with the Lindhard function replaced by the simpler collective approximation (see \eqref{eq:chi_CA}\,).

As mentioned above, employing rule 2 may appear cumbersome. 
This is avoided by neglecting $t_{\scriptscriptstyle\rm JF}^{\scriptscriptstyle\rm (3b)}$, which leads to a simplified (approximate) form of rule 2:
In turn, each loop formed by two exchange lines is replaced with $\frac{\hbar^2}{4m}\nabla^2\ell(r)^2$, and where more than two such lines make a loop, each exchange line, in turn, is replaced by $\frac{\hbar^2}{4m}\nabla^2\ell(r)$. Within this approximation we obtain
 \begin{align}
 g_{\scriptscriptstyle\mathrm{ee}}'\approx\>& 2\left[\ell + \nu (N_{\scriptscriptstyle\mathrm{cc}}\!+E_{\scriptscriptstyle\mathrm{cc}}) \right] (N'_{\scriptscriptstyle\mathrm{cc}}\!+E'_{\scriptscriptstyle\mathrm{cc}}) \nonumber \\
 &+ (N_{\scriptscriptstyle\mathrm{cc}}+E_{\scriptscriptstyle\mathrm{cc}})\frac{\hbar^2}{m}\nabla^2 \ell
+ N'_{\scriptscriptstyle\mathrm{ee}}+E'_{\scriptscriptstyle\mathrm{ee}}  \nonumber \\
&+ 2\left[ 1+N_{\scriptscriptstyle\mathrm{de}}\!+E_{\scriptscriptstyle\mathrm{de}} \right](N'_{\scriptscriptstyle\mathrm{de}}\!+E'_{\scriptscriptstyle\mathrm{de}}) \;.
\end{align}
Together with some specific approximation for the elementary diagrams, this provides an explicit expression for \(V_{\scriptscriptstyle\mathrm{ee}}\) in \eqref{eq: eff Veer}, if the ladder$^+$ approach is taken a step further.

\section{Derivation of $w_{\mathrm I}$ from PT}
\label{app: wI}
 
Here, we demonstrate how the induced potential \(w_{\mathrm I}\) is obtained from the ring (= bubble) diagrams of PT. As a first step the Lindhard function is approximated in the `collective'
(or `single pole') approximation, 
\begin{equation}
 \label{eq:chi_CA}
 \chi^{\scriptscriptstyle\mathrm{CA}}_0(q,\omega)\,=\frac{2t(q)}{(\hbar\omega +i\eta)^2- \big[t(q)/\SF(q)\big]^2} \;,
\end{equation}
implying an RPA response of Bijl--Feynman form,
 \begin{equation}\label{eq: RPA_CA}
 \chi_{\scriptscriptstyle\mathrm{RPA}}^{\scriptscriptstyle\mathrm{CA}}(q,\omega)
 \>=\>\frac{\chi^{\scriptscriptstyle\mathrm{CA}}_0(q,\omega)}
           {1- \tilde v(q)\, \chi^{\scriptscriptstyle\mathrm{CA}}_0(q,\omega) } \,.
 \end{equation}
This approximation will elucidate the physical meaning of \(w_{\mathrm I}\).
We next add, in a similar spirit, a two-point approximation of the ladder diagrams to the bare interaction, \(v\to v+V_{\mathrm{lad}}\equiv\!\Vph\). Note that the rungs in $V_{\mathrm{lad}}$ are not just bare, but rather a consistently resummed interaction, justifying the identification with the particle--hole irreducible diagrams  
(no specific \(V_{\mathrm{lad}}\) needs to be assumed here).
The  $m_0$ sum rule (or adiabatic connection) relates \(\Vph\) to the static structure: 
\begin{subequations}
\begin{align}
 S(q)=& -\!\!\int_0^{\infty}\!\!d\hbar\omega \; 
 \Im m\big[\chi_{\scriptscriptstyle\mathrm{cRPA}}^{\scriptscriptstyle\mathrm{CA}}(q,\omega)\big] \>=\,\frac{t(q)}{\epsilon(q)} \;,
 \\
 &\epsilon(q)^2 \,\equiv\> \frac{t(q)^2}{\SF(q)^2} \,+\, 2t(q)\,\widetilde V_{\!\scriptscriptstyle\mathrm{ph}}(q) \;.
\end{align}\label{eq: m0SR}\end{subequations}
This is identical with the sFHNC EL equation \eqref{eq: EL_q}.

The ring diagrams generated from  \(\widetilde V_{\!\scriptscriptstyle\mathrm{ph}}(q)\) are  
\begin{equation}\label{eq: Vind}
 \widetilde V_{\mathrm{ring}}(q,\omega) \>\equiv\> 
 \frac{\widetilde V_{\!\scriptscriptstyle\mathrm{ph}\!}(q)^2\,\chi^{\scriptscriptstyle\mathrm{CA}}_0(q,\omega)} {1-\widetilde V_{\!\scriptscriptstyle\mathrm{ph}\!}(q)\,\chi^{\scriptscriptstyle\mathrm{CA}}_0(q,\omega)} \;.
\end{equation}
Even though  \(\widetilde V_{\!\scriptscriptstyle\mathrm{ph}\!}(q)\) is static, this summation is energy dependent. Following Jackson et al.\,\cite{Jackson1985} we replace \(\omega\) in \(V_{\mathrm{ring}}\) by a suitably determined characteristic frequency $\bar\omega_q$. Again, the solely statistical effects are split off,
\begin{align}
 \chi(q,\omega) &=\> 
  \> \chi_0(q,\omega)  \>+\> \chi_{0}^2(q,\omega)
  \,\big[\,\widetilde V_{\!\scriptscriptstyle\mathrm{ph}\!}(q)+
           \widetilde V _{\mathrm{ring}}(q,\omega) \big]_{\phantom{|}}
  \nonumber\\
 &\approx\chi^{\scriptscriptstyle\mathrm{CA}}_0(q,\omega)+
         \chi^{\scriptscriptstyle\mathrm{CA}}_{0}(q,\omega)^2
         \big[\,\widetilde V_{\!\scriptscriptstyle\mathrm{ph}\!}(q)+
                \widetilde V _{\mathrm{ring}}(q,\bar\omega_q) \big]_{\phantom{|}} \>.\;
\end{align}
Demanding consistency with the static structure implies
\begin{subequations}\begin{equation}
 S(q) \>=\> S_{\scriptscriptstyle\mathrm{F}}(q) - 
 \frac{S_{\scriptscriptstyle\mathrm{F}}(q)^3}{t(q)}\,
 \frac{\widetilde V_{\!\scriptscriptstyle\mathrm{ph}\!}(q)}
      {1-\widetilde V_{\!\scriptscriptstyle\mathrm{ph}\!}(q)\,\chi^{\scriptscriptstyle\mathrm{CA}}_0(q,\bar\omega_q)}  \;.
\end{equation}
This results in (we skip \(q\) for ease of reading)
\begin{equation}\label{eq: Pi0CA wbar} 
 \widetilde V_{\!\scriptscriptstyle\mathrm{ph}}\,\chi^{\scriptscriptstyle\mathrm{CA}}_0(\bar\omega_q) \,=\,
 1 - \frac{\SF}{2S}\,\left[\frac{\SF}{S}\!+\!1\right] \;;
\end{equation}
solving for \(\bar\omega_q\) using \eqref{eq: m0SR} gives
\begin{equation}
 \hbar^2 \bar\omega_q^2 \,=\, 
 -\frac{\epsilon(q)\, t(q)^2/S^2_{\scriptscriptstyle\mathrm{F}}(q)}{\epsilon(q)+2t(q)/S^2_{\scriptscriptstyle\mathrm{F}}(q))} \;,
\end{equation}\end{subequations}
the fermion analog of the boson \(\bar\omega_q\) found in Ref.\,\cite{Jackson1985}. 

Inserting \eqref{eq: Pi0CA wbar} into \eqref{eq: Vind} we thus have shown that
\begin{equation}
 \widetilde w_{\scriptscriptstyle\mathrm I}(q) = 
 \widetilde V_{\scriptscriptstyle\mathrm{ring}}(q,\bar \omega_q) \;,
\end{equation}
i.e. \(w_{\scriptscriptstyle\mathrm I}\) in eq. \eqref{eq: eff interactionFHNC wI} is indeed the sum of  approximated ring diagrams with the effective interaction \(v+V_{\rm lad}\,\).
 
This procedure also explains the name `induced potential' for  \(w_{\scriptscriptstyle\mathrm I}\).
Defined as the difference between the screened and the bare interaction, \(\tilde v_{\mathrm{ind}\,}\) in plain RPA is
\begin{equation}
 \tilde v_{\mathrm{ind}} \,=\, \frac{\tilde v}{1-\tilde v\,\chi_0} - \tilde v \,=\,
    \frac{\tilde v\,\chi_0\, \tilde v}{1-\tilde v\,\chi_0} \;.
\end{equation}
Replacing the bare interaction by the particle--hole irreducible interaction demonstrates the connection.
 
\section{Implementation \label{app: Implementation}}
 
We here show a favorable self-consistent procedure to treat boson-like FHNC equations. Instead of numerically solving the non-linear differential equation (\ref{eq: realspace eqFHNC}), a straightforward manipulation maps it onto an algebraic relation in reciprocal space. The static structure factor defines an auxiliary potential, where we split off the boson induced interaction,
\begin{subequations}\begin{align}
 S(q) &\equiv\>1/\sqrt{1+\textstyle\frac2{t(q)}\displaystyle \widetilde V_{\rm aux}(q) }_{\phantom{\big|}} 
 \;, \label{eq: S from Vaux}\\
 \widetilde{w}_{\scriptscriptstyle\mathrm{IB}}(q) &\equiv\>
 -\frac{t(q)}2\,\Big[\,\frac1{S(q)}-1\Big]^2\,\big[2S(q)+1\big]_{\phantom{\Big]\big|\!\!\!\!}} 
 \;, \label{eq: wIB from S}\\
 \widetilde V_{\mathrm{aux}}(q) &=\> -\widetilde{w}_{\scriptscriptstyle\mathrm{IB}}(q) \,-\, t(q)\,\big(S(q)-\!1\big)\;.
\label{eq: Vaux q}\end{align}\end{subequations}
In the exact EL equation multiplied with \(\sqrt{g}\), where (as in Sec.\,\ref{ssec: relation to PT})
we subsume all effective interactions in \(V\) 
\begin{align}
 \Big[-\textstyle\frac{\hbar^2}{m}\displaystyle\nabla^2 \,+ &\,V(r) 
 \Big]_{\phantom{\big|}\!\!} g(r) \,+\, \textstyle\frac{\hbar^2}{m}\displaystyle \Big|\nabla \sqrt{g(r)}\Big|^2 \>=\, 0 \;
\end{align}
the term \(Vg\) is recognized as \(\,-V_{\mathrm{aux}\!}-w_{\scriptscriptstyle\mathrm{IB}}\,\), so that the auxiliary potential in real space is
\begin{align}
 V_{\rm aux}(r) \,=\> V(r)\,g(r)  
 \,-\,  w_{\scriptscriptstyle\mathrm{IB}}(r)  \,+\, 
 \textstyle\frac{\hbar^2}{m}\displaystyle \Big|\nabla \sqrt{g(r)}\Big|^2 \;.
\label{eq: Vaux r exact}\end{align}

In the ladder$^+$ approximation the explicit expression reads
\begin{align}
 V^{\rm lad^+\!}_{\rm aux}(r) \,=\> &\left[ 
 v(r)+w_{\mathrm I}(r)  +\, 
 \frac{\hbar^2\nabla^2\sqrt{g_{\scriptscriptstyle\mathrm F}(r)}}{m\sqrt{g_{\scriptscriptstyle\mathrm F}(r)}}
 \,\right]g(r) \nonumber \\
 &- w_{\scriptscriptstyle\mathrm{IB}}(r) \,+\, 
  \frac{\hbar^2}{m} \Big|\nabla \sqrt{g(r)}\Big|^2 \;.
\label{eq: Vaux r ladplus}\end{align}
An initial guess for \(g(r)\) gives  \(V_{\rm aux}(r)\) from \eqref{eq: Vaux r ladplus} and in turn, after a Fourier transform (FT) with the convention
\begin{equation}
 \widetilde V_{\rm aux}(q) \>=\> \rho\!\int\!d^3r\>  e^{-i{\bf q\cdot r}}\, V_{\rm aux}(r)
\label{eq: FT Vrfromq}\end{equation}
a static structure factor from \eqref{eq: S from Vaux}, inserted into \eqref{eq: wIB from S}. 
The inverse FT gives \(w_{\mathrm{IB}}(r)\) and a new \(g(r)\) from 
\begin{equation}
 g(r)-\!1 \>= \int\!\frac{d^3q}{(2\pi)^3\rho}\> e^{i{\bf q\cdot r}}\, \big(S(q)\!-\!1\big) \;.
\label{eq: FT Stog}\end{equation}\label{eqs: FT}
This procedure is iterated until convergence is achieved, merely having the FT as rate limiting step. Note that in contrast to STLS-type \cite{Singwi1969} methods, all FHNC-EL approaches presented here are easier to implement, since no integrations in addition to the FT are involved.

For the bFHNC of Kallio we adopt the same strategy. The auxiliary potential acquires an additional term \(w_{\mathrm{IBF}}\), which is determined solely by 
\(\SF\)
\begin{align}
 V^{\mathrm{Kallio}}_{\rm aux}(r)\,=\> &\Big[ v(r)+w_{\scriptscriptstyle\mathrm{IB}}(r)-w_{\scriptscriptstyle\mathrm{IBF}}(r)  +\,
 V_{\scriptscriptstyle\mathrm F}(r)
 \Big]\,g(r) \nonumber \\
 &- w_{\scriptscriptstyle\mathrm{IB}}(r)+ \frac{\hbar^2}{m}\Big|\nabla \sqrt{g(r)}\Big|^2 \;,
 \nonumber \\
 \widetilde{w}_{\scriptscriptstyle\mathrm{IBF}}(q)\,\equiv&
 -\frac12t(q)\,\Big[\,\frac1{S_{\scriptscriptstyle\mathrm F}(q)}-1\Big]^2\big[2S_{\scriptscriptstyle\mathrm F}(q)+1\big]  \;.
\label{eq: Vaux rK}\end{align}
This corresponds to approximating the exact real-space EL equation \eqref{eq: realspace eqFHNC} with
\begin{equation}
  \bigg[-\frac{\hbar^2}{m}\nabla^2 + v + w_{\scriptscriptstyle\mathrm{IB}}-w_{\scriptscriptstyle\mathrm{IBF}} + 
  V_{\scriptscriptstyle\mathrm F}\, \bigg]_{\phantom{\big|}\!\!} \sqrt{g(r)} \>=\, 0 \;.
\label{eq: Kallio eqFHNC}
\end{equation}
For \(v\!\to\!0\), the iterations yield \(w_{\scriptscriptstyle\mathrm{IB}}\to w_{\scriptscriptstyle\mathrm{IBF}\,}\). An appealing feature is that this concept can be easily extended to include elementary diagrams.

An implementation of all three versions discussed in the present work, the sFHNC, the bFHNC and the ladder$^+$ method can be found at \url{https://github.com/mpanho/FHNC_3D.git}.

\bibliographystyle{apsrev4-1} 
\bibliography{lib}
\end{document}